\documentclass{elsart}

\usepackage{pst-node,amssymb,hyperref,cite,enumerate}

\usepackage[utf8]{inputenc}

\newcommand{\ket}[1]{\ensuremath{|#1\rangle}}

\newcommand{\NZ}{{\field{N}}_0}
\newcommand{\field}[1]{\mathbb{#1}}

\newcommand{\rdirac}[1]{\mbox{$|#1\rangle$}}

\newtheorem{definition}{Definition}
\newtheorem{conjecture}{Conjecture}
\newtheorem{theorem}{Theorem}

\newcommand{\opm}[1]{}

\begin{document}

\opm{Nuwerige titel en in TCS-styl}

\begin{frontmatter}
\title{Universality and programmability of quantum computers}

\author{Willem Fouch\'e}
\address{Department of Decision Sciences}
\author{Johannes Heidema}
\address{Department of Mathematical Sciences}
\author{Glyn Jones}
\address{Department of Physics}
\author{Petrus H. Potgieter}
\address{Department of Decision Sciences}
\ead{%
php@member.ams.org, potgiph@unisa.ac.za
}
\address{University of South Africa, PO Box 392, Unisa, 0003, Pretoria}

\begin{abstract}
\opm{Ons moet hier ook maar mooi deurlees en o.a. kyk dat dit genoeg verskil van ons kort poging.}
Manin, Feynman, and Deutsch have viewed quantum computing as a kind of universal physical simulation procedure. Much of the writing about quantum logic circuits and quantum Turing machines has shown how these machines can simulate an arbitrary unitary transformation on a finite number of qubits. The problem of universality has been addressed most famously in a paper by Deutsch, and later by Bernstein and Vazirani as well as Kitaev and Solovay. The quantum logic circuit model, developed by Feynman and Deutsch, has been more prominent in the research literature than Deutsch's quantum Turing machines. Quantum Turing machines form a class closely related to deterministic and probabilistic Turing machines and one might hope to find a universal machine in this class. A universal machine is the basis of a notion of programmability. 
The extent to which universality has in fact been established by the pioneers in the field is examined and this key notion in theoretical computer science is scrutinised in quantum computing by distinguishing various connotations and concomitant results and problems.
\end{abstract}


\end{frontmatter}

\section{Introduction}

In its attempt to cast light on the concept of programmability in quantum computing, this paper discusses
two 
notions of ``universality'', namely 
\begin{itemize}
\item a universal set of generating ``components'' for a given class of machines, and
\item a universal machine in a class. 
\end{itemize} 
In his classic introduction \cite{Hermes1969} to the theory of recursive functions, Hans Hermes devotes Chapter 2 to the ``engineering'' of Turing machines (defined below).  He introduces a finite set of elementary Turing machines 
and %
then describes how to combine them to build more complex ones and proves that the elementary machines constitute a universal generating set: any Turing machine 
whatsoever is equivalent (in terms of input-output behaviour) to a Turing machine constructed by combining elementary machines. Then, in the last chapter of the book, Hermes proves that there exists a universal Turing machine which can simulate the input-output behaviour of any Turing machine  %
if provided with an appropriate program. 
These are instances of the two different notions of universality. 
Confusing and conflating the existence of a universal generating set with the existence of a universal machine can engender some conceptual perplexity, from which the literature on quantum computing does not seem to escape completely.  This contribution examines these issues in more detail.

\section{Classical and probabilistic Turing machines}

By the beginning of the twentieth century mathematicians had become quite interested in establishing a formal model of computability. In 1936 Alan Turing described an abstract device, now called a \textit{Turing machine}, which follows a simple, finite set of rules in a predictable fashion to transform finite strings (input) into finite strings (output, where defined). The Turing machine (TM) can be imagined to be a small device running on a two-way infinite tape with discrete cells, each cell containing only the symbol \textbf{0} or \textbf{1} or a blank. The TM has a finite set of possible internal states and a movable head that can read the contents of the cell of the tape immediately under it. The head may also, at each step, write a symbol to the cell over which it finds itself. There are two special internal states: an \textit{initial state} $q_0$ and a \textit{halting state} $q_H$.

A TM has a finite list of instructions, or \textit{transition rules}, describing its operation. There is at most one transition rule for each combination of cell content (under the head) and internal state. If the internal state is $q_{i}$ and the head is over a cell with content $S_{j}$ then the machine looks for a rule corresponding to $(q_i, S_j)$. If no rule is found, the machine enters the halting state immediately. If a rule corresponding to $(q_i, S_j)$ is found, it will tell the machine what to write to the cell under the head, whether to move one cell left or right or to stay put and which internal state to enter. There is no transition rule corresponding to the halting state. Sometimes we refer to the entire collection of individual rules for all the different $(q_i, S_j)$ as \textit{the transition rule} of the machine.

A \textit{computation} consists of starting the TM with the head over the first non-blank cell (which we may label position 0 on the tape) from the left of the tape (it is assumed that there is nothing but some finite \textit{input} on the tape) and the machine in internal state $q_0$. Now the transition rules are simply applied until the machine enters the halting state $q_H$, at which point the content of the tape will be the \textit{output} of the computation. If, for some input, the machine never halts then the output corresponding to that input is simply undefined. It is clear how every TM defines a (possibly, partial) function $f:\NZ\rightarrow\NZ$  from the set of counting numbers to itself.

Turing machines are the canonical models of computing devices. No deterministic device, operating by finite (but possibly unbounded) means has been shown to be able to compute functions not computable by a Turing machine. In fact, one may view one's desktop computer as a Turing machine with a \textit{finite} tape.

A probabilistic Turing machine (PTM) is identical to an ordinary Turing machine except for the fact that at each machine configuration $\left(q_i,S_j\right)$ there is a finite set of transition rules (each with an associated probability) that apply and that a random choice determines which rule to apply. We fix some threshold probability greater than even odds (say, 75\%) and say that a specific PTM computes $f(x)$ on input $x$ if and only if it halts with $f(x)$ as output with probability greater than  75\%. 

\section{Quantum Turing Machines (QTMs)}

A natural model for quantum computation is based on the classical Turing machine. The \textit{quantum Turing machine} (QTM) was first\footnote{
Paul Benioff had related a similar idea somewhat earlier \cite{Ben80a} but primarily in connection with presenting a possible physical basis for reversible computing.} 
described by David Deutsch \cite{Deu85a}.  The basic idea is quite simple, a QTM being roughly a probabilistic Turing machine (PTM) with complex transition amplitudes (the squared moduli of which add up to one at each application) instead of real probabilities. 

Without loss of generality everything can be assumed to be coded in binary so that each position on the tape of the QTM will 
correspond to a single qubit (quantum bit).  A unit of quantum information, the qubit is a two level quantum mechanical system, whose state is described by a linear superposition of two basis quantum states, often labelled \ket{0} and \ket{1}.
The actual (quantum) state space of the machine will be a direct sum of $n$-qubit spaces (where $n$ is an indication of how much tape has been used, each $n$-qubit space being the $n$-fold tensor of the single qubit space). 
The direct sum is, however, not a complete inner-product space and therefore---by the postulates of quantum mechanics---not a valid state space. However, the underlying Hilbert space can be taken to be the completion of the direct sum and a unitary operator $U$ on the direct sum (see \cite{BV97a}) can be extended to a unitary operator $\widehat{U}$ on the Hilbert space. This completed space and operator will correspond to the physical system associated with the QTM, thereby taking care of the \textit{physicality} of the QTM.

\subsection{Operation of a QTM}

\label{sec:def}

In the following the \textit{classical machine} is a machine with a two-way infinite tape, starting over position 0 on the tape as described above, that we use as a kind of template for the {quantum Turing machine}. The  corresponding {quantum Turing machine} (QTM) might work as follows (based on the {Deutsch} description \cite{Deu85a}, Ozawa \cite{Oza97a}, Bernstein and Vazirani \cite{BV97a}).
\begin{enumerate}[I.]
\item The quantum state space of the machine is spanned by a basis (here called the \textit{computation basis}) consisting of states 
$$\ket{h}\ket{q_C}\ket{T_C}\ket{x_C}$$ 
where $\ket{h}$ is the halt qubit, $h\in\{0,1\}$ and $(q_C,T_C,x_C)$ is a configuration of the corresponding classical machine, where $x_C$ denotes the position of the head, $q_C$ the internal state of the machine and $T_C$ the non-blank content of the tape. 
\item Special initial and terminal internal states have been identified (corresponding to the initial state and halting state of the classical machine).
\item The single  transition rule is now a unitary operator $U$ which, in each step, maps each basic $\ket{h}\ket{q}\ket{T}\ket{x}$ to a superposition of only finitely many  $\ket{h'}\ket{q'}\ket{T'}\ket{x'}$, where 
\begin{enumerate}
\item the rule is identical for $\ket{h}\ket{q}\ket{T_1}\ket{x}$ and $\ket{h}\ket{q}\ket{T_2}\ket{y}$ when $T_1$ in position $x$ and $T_2$  in position $y$ have the same content, i.e. the rule depends only on the content of the tape under the head and the internal state $q$ and not on the position of the head or on the content of the rest of the tape;
\item $T'$ and $T$ differ at most in position $x$;
\item $|x'-x|\leq 1$ (depending on whether the corresponding classical machine moves one position to the left, to the right, or not at all);
\item $h'=1$ if and only if $q'$ is the halting state of the classical machine; and
\item $T'=T$, $q'=q$ and $h'=h$ whenever $h=1$.
\end{enumerate} 
Finitely many subrules
\begin{equation}
\label{eqtr}
\ket{h}\ket{q}\ket{T}\ket{x} \quad \longmapsto \quad \sum_{i=1}^n c_i \ket{h_i}\ket{q_i}\ket{T_i}\ket{x_i}
\end{equation}
will determine $U$ as there are, by the stipulations above, only finitely many possible---given that the alphabet of the tape (binary in our case) and the number of internal states are both finite.
Note that the transitional rule (``program'') will have a finite specification only if the transition amplitudes in the superposition of the $\ket{h'}\ket{q'}\ket{T'}\ket{x'}$ are all \textit{computable} complex numbers, which we will of course assume to be the case throughout. The transition rule can also, obviously, be extended (linearly) to finite superpositions of $\ket{h}\ket{q}\ket{T}\ket{x}$.
\item The machine is started with a finite superposition of inputs
$$\ket{0}\ket{q_0}\ket{T}\ket{x}.$$
Because of the form that the transition rule is  allowed to take (and the fact that there are only finitely many internal machine states) the machine will be in the superposition of  only \textit{finitely many} basic states $\ket{h}\ket{q}\ket{T}\ket{x}$ at any step during the entire run\footnote{A more hazy concept than for classical Turing machines, as a QTM only really stops when one has observed the halt qubit and the content of the tape, so one may think of the transition rule being applied \textit{ad infinitum}, step-by-step, unless the operator (physically, classically and externally) stops the machine.} of  computation.
\end{enumerate} 
The description of the machine given here differs from a classical reversible Turing machine in two obvious respects.
\begin{enumerate}[(a)]
\item Transition rules are allowed to map a state of the machine to the superposition of several states. The crucial distinction with classical probabilistic machines is that the QTM goes to a quantum superposition of states whereas the classical PTM can be seen as either going to a classical probability distribution over states or to a specific state with some classical probability. Quantum computing, of course, uses superposition in an essential way\footnote{A very readable and accessible explanation of how and why this works can be found in \cite{fortnow_one_2000}.}---as in the algorithms of Shor or Grover.
\item The input is allowed to be a superposition of a finite number of ``classical'' inputs.
\end{enumerate}
It is not immediately obvious why a finite collection of specifications of the form (\ref{eqtr}) should necessarily define a unitary $U$, however, just as it might not be apparent why a finite collection of rules 
$$\ket{h}\ket{q}\ket{T}\ket{x} \quad \longmapsto \quad \ket{h'}\ket{q'}\ket{T'}\ket{x'}$$
for a classical machine would necessarily specify a reversible machine. Unitarity is, of course, a precondition for the quantum device to be feasible.

We demonstrate that the specification of what QTMs are, is at least consistent, that such beasties exist mathematically. Let the underlying classical template be a reversible TM, which, after reaching its halting state, keeps moving its head stepwise in one direction without ever changing anything on its tape. (Note that III.(e) does not stipulate that $x'=x$ whenever $h=1$; to be reversible, something must change at every step.) The corresponding QTM is now constructed by linearly generating its $U$ according to the corresponding transition rules of the simple form above --- with no superpositions. Then this $U$ is determined by a permutation of the computation basis of the QTM, and hence is unitary. Should the rules involve superpositions (as happens in all interesting cases), a proof of the unitarity of the induced $U$ is called for.

\subsection{Time evolution of the QTM and halting}

If $U$ is the operator that describes one application of the transition rule (i.e. one step in the operation) of the machine, then the evolution of an  unobserved machine (where not even the halt bit is measured) for $n$ steps is simply described by $V=U^n$. If the first measurement occurs after $n_1$ steps, and the measurement is described by an operator $J_1$, then the evolution of the machine for the first $n_1+j$ steps is described by $$U^jJ_1U^{n_1}$$ which is in general no longer unitary since the operator $J_1$ is a measurement (always in the computational basis). It is important to note that the machine evolves unitarily only when no measurement takes place at all. 

The output of the machine is on the tape as a superposition of basis states and should be read off after having measured the content of the halt bit and finding it in the state 1. The operator may at any time measure the halt bit\footnote{
The halt \textit{qubit}, of course, until we measure it.} in order to decide whether to read the tape content (and collapse the state of the machine to one of the basis states). The halt bit is intended to give the operator of the machine an indication of when an output may be read off from the tape (and by observation collapsing the system to an eigenstate) without interfering excessively with the computation. It seems that Deutsch's original idea was that there would be no entanglement at all between the halt bit and the rest of the machine, but this cannot be guaranteed. 
The \textit{output} of a QTM for some specific input $x$ (which may be a superposition of classical inputs) is a probability distribution $P_x$ over all possible contents of the tape at the time of observing the halt bit to have been activated. Note that the observation of an activated halt bit may in itself be a random event, but it has been argued by Ozawa \cite{Oza97a} and others, that $P_x$ does not depend on the random \textit{observation} events.


\section{Universality and programmability in the machine model}

The notion of a \textit{universal} computing device in a specific class is crucial for the development of a complexity theory and---more basically---establishes the notion of programmability. Naturally, we will start the discussion by reviewing the well-established notions of universality in classical deterministic and probabilistic computing before moving on to examine the concept of a ``universal QTM'' introduced by Deutsch.

\subsection{Classical universality and programmability}

Consider a general countable class of machines, say \textit{Manchester machines } (MMs), that compute partial functions, i.e. functions that are not necessarily defined for all inputs (since the machine might not halt, for example, as in the case of a Turing machine). 
Since there are only countably many machine descriptions, let us assume that each Manchester\footnote{Alan Turing worked on building and programming one of the first electronic computers in the city of Manchester after the Second World War.} 
machine is fully described by a natural number. It should be possible to recover the full description of the machine's functioning from the natural
number in an effective way, so it should not simply be any enumeration of the countable set. Let $\Phi_n$ denote the partial function computed by machine $n$ and fix an MM-computable bijective function\footnote{%
It will strike the attentive reader that $h$ is the first (and last) function of two variables to appear here but that we have implicitly considered Manchester machines with one natural number input only. This is illogical, but the problem can be fixed in a well-established way. Suffice it to say that one should be able to consider $h$ MM-computable in an obvious and logical way. One only needs \textit{one} such function $h$ and we will therefore not elaborate here.
} $h : \NZ \times \NZ \rightarrow \NZ$,
assuming such a function exists\footnote{If it does not, the class of machines would really be very poor. It would not make a big difference if we took, for example, $h:(x,y)\mapsto2^x3^y$ instead of an onto function but the convention that $h$ be onto is harmless and convenient.}.
\begin{definition}
\label{defone}
  If there exists a number $N$ such that
  \[ \Phi_N \left( h ( n, m ) \right) = \Phi_n ( m ) \]
  which means that the functions are either equal and both defined or both
  undefined, for all $n$ and $m$, then the machine described by $N$ is called a
  \textit{Universal Manchester Machine} (UMM).
\end{definition}
Programmability is firmly linked to the concept of universality and is, of course, a necessary condition for universality.  Is it a sufficient condition? 
A particular Turing machine is usually thought of as dedicated to a particular task, defined by a set of quintuples describing the operations to be carried out in  sequence.  Every Turing machine has thus a finite description (of internal states, tape entries and operation rules---which are unbounded but finitely many) which could be used as input to another Turing machine.
 
A universal Turing machine, (of which there are infinitely many), can simulate all the Turing machines, and is thus  programmable for the entire class of Turing 
machines.  If a machine is programmable for any device in its class, then it is universal. 
Not all programmable computer devices are universal in any  sense. In fact, one could use the term programmable to describe any device taking input of the form $\langle p,x\rangle$ where $p$ is the ``program'' and $x$ the ``data'' and where the action of the machine on $x$ depends on $p$.  Such machines are universal (for their class) only if they can---through the suitable choice of $p$---mimic the operation of any other machine in the class.

\subsection{Probabilistic Turing machine universality}

Since halting is a probabilistic notion for a QTM, the notion of universality for quantum devices should be akin to that for probabilistic machines. For probabilistic machines, however, Definition \ref{defone} does not directly apply and it is necessary to generalise it as follows.

\begin{definition}
\label{deftwo}
 If there exists a number $N$ such that
 \[ \Phi_N \left( h ( n, m ) \right) = \Phi_n ( m ) \]
 which means that the functions are either equal and both defined or both
 undefined \textbf{(if deterministic) and if not deterministic then the values have the same distribution}, for all $n$ and $m$, then the machine described by $N$ is called a \textit{Universal Manchester Machine} (UMM).
\end{definition}

In the case, for example, of deterministic Turing machines (which are a strict subset of the probabilistic machines) the two notions of universality coincide, of course. The main aim of this section is to discuss this (second) notion of universality for quantum Turing machines (QTMs).
 
One can easily show, incidentally, that every function $f$ which can be computed in this sense by a PTM, is also computable by some ordinary TM in the usual sense. Nevertheless PTMs have always been of interest because probabilistic algorithms can often be found that are quite fast by comparison to the best known classical procedure. The class of PTMs is often defined by restricting the probabilities to $\frac{1}{2}$ or $1$ only. In this case the class can also be obtained by taking the ordinary TMs and adding a special write instruction to write one random bit to the tape. 
The PTMs are often described, in this model, as ``TMs with access to a fair coin toss''. It is easy to see how a universal machine might be described in this class: it would simply be a universal TM equipped with the random output instruction. Such a \textit{universal PTM} (UPTM) could obviously simulate any other ``coin toss'' PTM perfectly, by which is meant that the output of the UPTM would have exactly the same distribution as the output of a PTM for which it is executing a program. 

Now, which PTMs should our UPTM be able to simulate exactly? Well, since each PTM should have a finite description, the UPTM need only be able to simulate a countable collection of PTMs. Let us restrict the set of PTMs to those with  \textit{computable} transition probabilities. Each such machine is fully described by the finite set of transition rules and programs for computing each of the associated probabilities. This description is finite---thanks to the restriction of the probabilities to computable numbers. 

Since there is no reasonable way of giving a finite description of PTMs with non-computable transition probabilities, apart from the usual paradoxical definitions  of the type ``one more than the largest number which can be described in thirteen words'', this concludes the discussion for PTMs. 
Introducing arbitrary real transition probabilities makes no sense as it would immediately make any subset of the natural numbers decidable by a probabilistic machine.

\subsection{A universal QTM?}

Deutsch introduced a ``universal quantum computer'' (uQC, where u has not been capitalised in order to emphasise the difference between this universality concept and the preceding) in \cite{Deu85a}. The Deutsch uQC is in effect a QTM as in Section 3, based on a classical UTM with some additional (8 in \cite{Deu85a}) operations that allow any unitary transformation on one qubit to be approximated arbitrarily closely. 
Deutsch showed in the paper that for any given $L$, $\varepsilon>0$ and quantum device $U$ operating on $L$ qubits,  there exists a program $p_L$  (a classical finite string of bits)  for the uQC that (with input \ket{p_L} followed by any finite superposition of $L$-qubit basic states) approximates the operation of $U$ on the finite superposition of $L$-qubit basic states with accuracy at least $\varepsilon$ (in the inner-product norm). 
This is not the same kind of universality that we have for probabilistic and for deterministic Turing machines and even the concatenation scheme used by Deutsch has been questioned (for example, by Shi  \cite{Shi2002}).

Now, if we consider the earlier (second) definition of universality, then there can be no universal machine for the simple reason that in Deutsch's scheme there are uncountably many (transition rules for) QTMs.  For broadly the same reasons as outlined above for PTMs, 
we shall restrict ourselves henceforth to QTMs with computable transition amplitudes, i.e. transition amplitudes for which both the real and imaginary parts are computable numbers. 
We now fix a scheme for encoding the QTMs and associate any machine $M$ with the smallest\footnote{%
Two distinct natural numbers may, of course, encode physically identical machines.
} natural number that encodes it. 
Note that we say that a QTM outputs $y$ with probability $p$ if the probability of \textit{ever} observing the machine to be in the halt state with the tape in state \ket{y} is $p$. 
Does a universal machine for the (restricted) class of QTMs in the sense of Definition \ref{deftwo} exist?

Deutsch provided the rather incomplete solution mentioned above. Bernstein and Vazirani \cite{BV97a} have given another partial solution. They showed that there exists a quantum Turing machine $\mathcal U$ (they actually wrote $\mathcal M$) such that  \cite{BV97a}
\begin{quote}
``for any well-formed\footnote{Meaning that the time evolution operator is unitary.} 
QTM $M$, any $\varepsilon>0$, and any 
$T$, $\mathcal U$ can simulate $M$ with accuracy $\varepsilon$ for $T$ steps with slowdown polynomial in $T$ and $\frac{1}{\varepsilon}$.''
\end{quote}
The slowdown and the program for $\mathcal U$ \emph{both depend here on the length of the input}. The full Bernstein-Vazirani result could be summarised by the statement that
\begin{quote}
there exists a QTM $\mathcal U$ such that  for each QTM $M$ with finite description $\bar{M}$, $n$, $\varepsilon$ and $T$ there is a program ${\mathcal P}(\bar{M},n,\varepsilon,T)$ and a function $f_{\bar{M}}(T,n,\frac{1}{\varepsilon})$  (both recursive in their inputs) such that running $\mathcal U$ on input $\ket{{\mathcal P}(\bar{M},n,\varepsilon,T)}\otimes\ket{x}$ where $|x|=n$ for $f_{\bar{M}}(T,n,\frac{1}{\varepsilon})$ steps results---within accuracy $\varepsilon$---in the same distribution over observable states as running $M$ on input $\ket{x}$ for $T$ steps.
\end{quote} 
The simulation is clearly \emph{only approximate}. What Bernstein and Vazirani mean ``with accuracy $\varepsilon$'' is that if $P$ is the probability distribution over all observable states of $\mathcal U$ after $f_{\bar{M}}(T,n,\frac{1}{\varepsilon})$  steps with the given input and $Q$ is the corresponding probability distribution of $M$ after $T$ steps then $$\frac{1}{2} \sum_x |P(x)-Q(x)| ~\leq~ \varepsilon$$
where the summation is over all possible observable states $x$. Again, approximate simulation is quite different from the universality concept for ordinary and for probabilistic Turing machines (with computable probabilities) as in the latter cases the universal machine's simulation was \textit{exact}. 
Running $\mathcal U$ for exactly  $f_{\bar{M}}(T,n,\frac{1}{\varepsilon})$ steps on any input $\ket{{\mathcal P}(\bar{M},n,\varepsilon,T)}\otimes\ket{x}$ will have simulated the running of $M$ on \ket{x} for $T$ steps. We may not let $\mathcal U$  
run for any more steps as the state of the machine might then drift away from the to-be-simulated state of $M$ after $T$ steps. 
This behaviour is rather different from that of the UTM or UPTM---where there is no need to restrict the number of steps executed! 

What about the input to the machine? In general, the input to a QTM is allowed to be a (finite) superposition of basis states of the tape but the Bernstein-Vazirani theorem quoted here applies to a single state only. This is not a problem: it is straightforward to see that it also applies to a superposition of $m$ basic states (just replacing $\varepsilon$ by $\frac{1}{m}\varepsilon$).

Now, the Bernstein-Vazirani machine $\mathcal U$ immediately suggests the following semi-universal hybrid device (SUHD). The device takes the description $\bar{M}$ of a QTM $M$ as well as $x$ and $\varepsilon$ (which may be taken to be rational) as input. The machine operates as follows.
\begin{quote}
\tt 
\begin{tabbing}
$T$:= 1;\\
$n$:= |$x$|;\\
do \= \\
\> compute $P$ := ${\mathcal P}(\bar{M},n,\frac{\varepsilon}{T},T)$;\\
\> compute $S$ := $f_{\bar{M}}(T,n,\frac{T}{\varepsilon})$;\\
\> run $\mathcal U$ on $\ket{P}\otimes\ket{x}$ for S steps;\\
\> signal that quantum part of device may be observed;\\
\> wait a little;\\
\> reset quantum part of device;\\
\> $T$:=$T$+1;\\
while true;\\
\end{tabbing}
\end{quote}
Note that by replacing $\varepsilon$ by $\frac{\varepsilon}{T}$ we have ensured that by simply letting the SUHD run, we will not only be able to observe the simulated behaviour of $M$ for ever longer times, but also with ever-increasing accuracy. However, the SUHD is still not universal for the class of QTMs in the sense of Definition \ref{defone} or Definition \ref{deftwo}. This is true not only for the very obvious reason that its simulation is only \textit{approximate}, but for the much more fundamental reason that we do not know whether it is a QTM itself!

The SUHD is a real hybrid device which consists of a classical Turing-type machine and a quantum part. The SUHD is---in a sense---a robot capable of operating a quantum device (which forms part of itself) and there is no reason to think that such a robot cannot be built. The problem lies therein that the robot only gives a signal when we might observe the quantum part of the device. It cannot know whether we have observed the quantum part or not---otherwise the observer would become part of the device...

Now, any quantum device operates reversibly. In the case of the SUHD the step ``reset quantum part of device'' is the part which can be problematic in this regard. If the quantum part was not observed during the step ``wait a little'' then the inverse of the evolution operator of $\mathcal U$ can be used to effect such a reset. But, what if the observer(s) did make an observation of the quantum part during ``wait a little''? Now, the  inverse of the evolution operator of $\mathcal U$ will \textit{not} ``reset quantum part of device''. This is really a serious problem. In an ordinary QTM the evolution of the machine continues even when the halt bit has been observed, but for the SUHD even the observation of the halt bit (which may be in a superposed state, although not necessarily entangled with the rest of the machine) renders the operation of the device non-reversible. This is simply because for the ordinary QTM, the evolution 
operator can continue after the halt bit has been observed without perturbing the probability distribution that has been defined to be the QTM's \textit{output} (according to Ozawa \cite{Oza97a}) since the observation projects one, in a certain sense, only into a specific ($h=0$ or $h=1$) branch of the computation. For the hybrid device it is not that simple since the resetting step requires an undisturbed quantum part. If the quantum part has been disturbed at $T=k$, the operation described above will not be able to correctly reset the quantum part of the device and will not execute the loop faithfully for $T=k+1$. 
Of course, it is always possible for the operator to be instructed to restart the hybrid device after observation, but then we will be dealing with a new kind of bio-hybrid device and not a universal machine at all. In classical computing this would be the equivalent of the user strictly having to reboot the computer each time after looking at the screen, i.e. there would be no \textit{autonomy} of operation. Pure quantum computing devices are prevented by  the No Cloning Theorem from copying initial configurations of substems, which precludes the realisation of such a na\"\i ve hybrid operation by a quantum device.

\begin{conjecture}
The SUHD derived from Bernstein and Vazirani's $\mathcal U$ cannot be made to operate reversibly and is therefore not a QTM.
\end{conjecture}
The immediate consequence of the conjecture is that (as yet) no universal machine has been shown to exist in quantum computing and that the notion of universal programmability has not really been established for quantum computing in the QTM model.
\opm{Opmerking hier oor Markus Mueller se poging.}

\section{Quantum gates}

Quantum gates provide another (and more practical) engineering paradigm for quantum computation initiated by Richard Feynman \cite{Fey86a} and David Deutsch \cite{Deu89a}. The classical analogue is a logic circuit. In principle, in the quantum gate model, a quantum computation works as
follows.
\begin{enumerate}
\item The first step typically involves the preprocessing of the input data on a {\it
    classical} computer. For example, in the Shor algorithm for the factoring
    problem we must ensure in a classical way that the input number is not
    a prime power.
\item Based on these preprocessed data, we have to prepare the quantum
  register. This means, in the simplest case, to prepare classical data e.g. a binary string $x$ of length $d$, say, as the state $\rdirac{x}$ in $2^d$-dimensional Hilbert space. In most cases, however, one would be required
  to prepare a superposition of states $\rdirac{x}$.
\item Next we apply the quantum circuit $C$, which is a sequence of local
  quantum operators,  to the input state $\rdirac{\phi}$
  and after the calculation we get the output state $U\rdirac{\phi}$ where $U$
  is the unitary operator corresponding to $C$.
\item To read out the data we perform a von Neumann measurement in the
  computational basis.
\item Finally we may have to post-process the value on a classical computer. In
  general we obtain a correct result with probability less than one, which means we
  have to check the validity of the result with a polynomial time algorithm
  and if wrong, we have to go back to step 2.
\end{enumerate}
Hence, in this model, a quantum computation is a hybrid of classical and
probabilistic algorithms coupled with quantum evolutions of prepared quantum states.

Suppose $V$ is a unitary operator acting on ${\mathcal H}^{\otimes f}$. For $d
\geq f$, we call a unitary operator on ${\mathcal H}^{\otimes d}$ an {\it instance} of $V$ if
it is any operator acting like  $V$ on a fixed $f$ of the possible $d$ qubits and as the identity on the remaining qubits. 
In order to discuss programmability in this context, we introduce the idea of instruction sets.
An {\it instruction set} $G$ for a multiqubit of a fixed length $d$ is a {\it finite} set
of quantum gates satisfying the following conditions.
\begin{itemize}
\item All gates $V \in G$ are in $SU(2^d)$, that is, they are unitary operators
  on the $2^d$-dimensional Hilbert space ${\mathcal H}^{\otimes d}$ where $\mathcal H$ is
  $2$-dimensional over $\field{C}$ and each operator has determinant one.
\item For each $V \in G$ the inverse operation $V^\dagger$ also belongs to
  $G$.
\item The group generated by $G$ is topologically dense in $SU(2^d)$. This means
  that for any given quantum gate $U \in SU(2^d)$ and any degree of accuracy
  $\epsilon > 0$, there exists a finite product $V=V_1V_2 \ldots V_k$ of
  instances of gates from $G$ which is an
  $\epsilon$-approximation to $U$, that is to say, such that $$||U-V_1V_2
  \ldots V_k|| < \epsilon.$$
\end{itemize}
Suppose $U$ and $V$ are two unitary operators on the same state space with $U$
the target unitary operator that we wish to implement and $V=V_1V_2 \cdots V_k$ being the unitary
operator that is actually implemented from an instruction set as above. Let $M$
be a positive operator valued measure (POVM) element associated with the
measurement and let $P_U$ (or $P_V$) be the
probability of obtaining the corresponding measurement outcome if the
operation $U$ (or $V$) is performed with a
starting state $\rdirac{\phi}$. Then it can be shown that (see \cite{NC00a}, p195)
\[|P_U-P_V| =|\langle\phi|U^\dagger MU|\phi\rangle -\langle\phi|V^\dagger MV|\phi\rangle| \leq 2||U-V||.\] 
This inequality gives quantitative expression to the idea that when the error
$||U-V||$ is small, the difference in probabilities between measurement
outcomes is also small.

An example of a set of universal gates is that ``generated'' by instances of $T$, the Toffoli gate, 
$H$, the Hadamard gate, and the phase gates. It is ``generated'' in the following sense: We consider
all unitary operators for $d$-multiqubits which is a product of instances
of  $H$, $T$, the phase gates together with their inverses. Then this set $G$ is
an instruction set for multiqubits of length $d$.

\section{Universal sets of gates: the Solovay-Kitaev Theorem}

The problem of quantum compilation is the following: Given an instruction set
$G$, how can we approximate an arbitrary quantum gate by means of a finite
sequence of instructions from $G$ in a manner which is both effective (i.e.,
computable in the classical sense), and efficient as far as both the time and
space complexity are concerned. The Solovay-Kitaev theorem \cite{KSV02a} is a truly
remarkable contribution to this problem:

\begin{theorem}
Let $G$ be an instruction set for $SU(2^d)$, and let a desired measure of
accuracy $\epsilon > 0$ be given. There is a universal constant $c$ such that
for any $U$ in $SU(2^d)$, there exists a finite sequence $S$ of gates from $G$ of
length $O_d\left(\log^c(1/\epsilon)\right)$ such that the product of the sequence $S$ is
within $\epsilon$ of $U$ with respect to the operator norm. 
\end{theorem}

More precisely, an arbitrary unitary operator $U$ on $d$ qubits can be
approximated to within a distance $\epsilon$ in the operator norm 
by using
$\mbox{O}\left( d^24^d\log^c(d^24^d/\epsilon)\right)$ instances of gates from $G$. 
This can be shown%
\footnote{%
The proof in \cite{NC00a}, pp199--200, is believed to contain an error. An upcoming paper, \textit{On the Solovay-Kitaev theorem}, by W.L. Fouché contains a full proof.%
} to be close to optimal in the following sense: For a given instruction set $G$ and a measure
of accuracy $\epsilon > 0$, there are unitary transformations $U$ on $d$
qubits which take $\Omega(2^d \log(1/\epsilon)/\log(d))$ instances of gates from $G$ to
implement an approximation $V$ such that $||U-V|| < \epsilon$. 

Many authors state that this can be done in an effective and efficient
manner. This must be read with some care! We call a unitary operation {\it recursive}
with respect to the chosen measurement basis if all its matrix entries relative to this basis are
recursive complex numbers. Recall that a complex number is a \textit{recursive} complex
number provided both its real and imaginary parts are recursive real
numbers. A real number $x$ is recursive if there is an algorithmic procedure which
with input a natural number $n$ will yield a binary rational number of the form $\ell/2^n$
such that $|x-\ell/2^n|<1/2^n$. Suppose now that all the matrix entries
of the gates in $G$ with respect to the orthonormal basis in which the
measurement is performed are recursive complex numbers, 
but that $U$ is not recursive relative to this basis. Suppose we have an
effective procedure that will yield for any given natural $n$ descriptions of
instances $V_1, \ldots, V_k$ of gates from $G$ such that
$||U-V_1 \cdots V_k||<1/n$. Then it is clear that all the matrix coefficients of
$U$ with respect to the measurement basis are recursive complex numbers---a contradiction. 
Our impression is that it may  be possible to effectively compute $U$ provided $U$ is recursive with respect to the measurement basis. Then, the time complexity of finding the sequence $S$ of gates will depend on the complexity of determining the elements of $U$ and of doing the required algebraic operations. 
The claim is that this accuracy can be obtained using
$O_d\left(\log^{2.71}(1/\epsilon)\right)$ computational steps. As we understand matters at
this stage, this is correct if the computation is relative to an {\it oracle} that has complete
information about  $U$ with respect to the measurement basis. It remains
to be investigated how the recursive complexity of $U$ affects this
claim. It is also
ambiguous to state that the approximation can be done in an efficient manner, 
for as stated  above, the worst case approximation will always be at least
exponential in the length of the multiqubit on which the unitary transformations operate.

\section{Conclusion}

Research into quantum computation over the past 20 years has been very successful in stimulating the development of quantum cryptography (already in industrial application), the study of quantum information and the discovery of novel algorithms for traditionally hard and interesting problems such as prime factorisation. One would be wise, however, to heed the words of Andrew Steane \cite{0034-4885-61-2-002}:
\begin{quote}
The title quantum computer will remain a misnomer for any experimental device
realised in the next twenty years. It is an abuse of language to call even a pocket calculator
a computer, because the word has come to be reserved for general-purpose machines
which more or less realise Turing's concept of the universal machine. The same ought to
be true for QCs if we do not want to mislead people. 
\end{quote} 
This paper has attempted to explain why certain (strong and interesting) results in quantum computation still fall short of establishing universality (and programmability) for quantum computing. At the very least, researchers in the field should attempt to explain how the results of Deutsch, Bernstein and Vazirani, Solovay, Kitaev and others can be used or expanded to construct a fully programmable universal quantum device. In the worst case, one needs to prove that such a fully universal quantum computer does not exist.
\opm{Miskien 'n opmerking hier oor die probleem van nabootsing van $U^n$ vir arbitrêr groot $n$ en dekoherensie.}
We conjecture that, like the QTMs and quantum circuits, other approaches to quantum computation which boast ``universal resources'' (e.g. measurement-based quantum computation \cite{nest2006}) may confront similar conceptual problems in the quest for a universal programmable machine.

\bibliographystyle{elsart-num}

\end{document}